\documentclass[conference]{IEEEtran}
\IEEEoverridecommandlockouts
\usepackage{cite}
\usepackage{amsmath,amssymb,amsfonts}
\usepackage{algorithmic}
\usepackage{graphicx}
\usepackage{textcomp}
\usepackage{xcolor}
\usepackage[utf8]{inputenc}
\usepackage{float}

\def\BibTeX{{\rm B\kern-.05em{\sc i\kern-.025em b}\kern-.08em
    T\kern-.1667em\lower.7ex\hbox{E}\kern-.125emX}}
\begin{document}

\title{What is the people posting about symptoms\\
related to Coronavirus in Bogota, Colombia}

\author{
\IEEEauthorblockN{Josimar Edinson Chire Saire}
\IEEEauthorblockA{Institute of Mathematics and\\
Computer Science (ICMC) \\
University of São Paulo (USP)\\
São Carlos, SP, Brazil\\
jecs89@usp.br}

\and
\IEEEauthorblockN{Roberto C. Navarro}
\IEEEauthorblockA{Faculty of Biossitemas\\ 
University of ABC\\
SP, Brazil\\
roberto.navarro@ufabc.br}
}

\maketitle

\begin{abstract}
During the last months, there is an increasing alarm about a new mutation of coronavirus, covid-19 coined by World Health Organization(WHO) with an impact in many areas: economy, health, politics and others. This situation was declared a pandemic by WHO, because of the fast expansion over many countries. At the same time, people is using Social Networks to express what they think, feel or experiment, so this people are Social Sensors and helps to analyze what is happening in their city. The objective of this paper is analyze the publications of Colombian people living in Bogota with a radius of 50 km using Text Mining techniques from symptomatology approach. The results support the understanding of the spread in Colombia related to symptoms of covid19.

\end{abstract}


\section{Introduction}
The impact of the epidemics of coronavirus  2019 (COVID-19) in a globalized world and with more communication tools allows instantaneous communication and in many cases without verification of the source of the information that it shows may have contraventions for society \cite{b1}.

For another place, the infoveillance for through the use Twitter (www.twitter.com) can be useful for longitudinal text mining and analysis to allow the analysis of some conditions of the epidemiology in real time as previously described in 2009 by Chew in the H1N1 pandemic \cite{b2}

Meanwhile, public health professionals have a increasingly need to establish a feedback loop and monitor real-time online public response and insights during emergency situations to examine the effectiveness of knowledge translation strategies and adapt future communications and educational campaigns to help the population face this pandemic \cite{b3}.

The dissemination of information can strongly influence people's behavior and alter the effectiveness of countermeasures implemented by governments. In this regard, models to predict the spread of the virus are beginning to monitor the behavioral response of the population with respect to public health interventions and the communication dynamics behind content consumption \cite{b4}.

During the last weeks, a big interest about Coronavirus started because of one infection located in Wu-han city in China,the epidemic scale of the recently emerged novel coronavirus  in Wuhan, China, has increased rapidly, with cases arising across China and other countries and regions. using a transmission model, it was  estimate of 81008 cases and the wuhan city have 21022 (11090-33490)  total infections in 1 to 22 January\cite{b5}. 

In relation to Colombia, the first case was registered in Bogota, Colombia. A girl of  19-year-old who returned to Bogota 26 February  from Milan, Italy. The woman was recovering at her place of residence. Before this, the young woman was placed in quarantine at her place of residence, with constant medical supervision, and after approximately 10 days it was confirmed that she had overcome the virus and was no longer infected with covid-19. The mayor pointed out that "contagion to her relatives was also avoided"\cite{b6}.
Regarding the incidence of COVID-19, it is estimated that by March 18, 2020 in Colombia there are 93 and 2 died according to the record of the Colombian Secretary of Health\cite{b7}.

The objective of this article is to describe the epidemiological impact of COVID-19 on press publications for 7 days before describing the first case of COVID-19 in Bogota, Colombia. With this, it is intended to describe the publications on twitter associated with the signs of the coronavirus with the advance of the pandemic and the persistence of the people of Bogota in this regard. This paper follows the next organization: section 2 explains the methodology for the experiments, section 3 presents results and analysis. Section 4 states the conclusions and section 5 introduces recommendations for studies related.


\section{Methodology}
The present work performs experiments with source data from Twitter with Natural Language Processing and Data Mining(Text Mining) following the next steps: 

\begin{itemize}
    \item Gather the relevant terms to search on Twitter
    \item Build the query for Twitter and collect data
    \item Pre-processing data to eliminate words with no relevance(stopwords)
    \item Visualization
\end{itemize}

\subsection{Gather Relevant Terms}
Following the next papers\cite{b8,b9} were extracted the next terms and translated to Spanish:

\begin{itemize}
\item 'fiebre','tos','gripe','estornudar','contagio','garganta'
\item 'dolor\_cabeza','dificultad\_respirar','congestion nasal','mialgia'
\item 'produccion\_esputo', 'hipoxemia', 'fatiga'
\end{itemize}

\subsection{Build the Query and collect data}

The extraction of tweets is through Twitter API, with the next parameters:
\begin{itemize}
    \item date: from 29-12-2019 to 14-03-2020
    \item terms: the words about symptoms in the previous subsection
    \item geolocalization: the capital of Colombia is Bogota(4.6,-74.083333)
    \item language: Spanish
    \item radius: around 50 km
\end{itemize}

\subsection{Preprocessing Data}
\begin{itemize}
    \item Change format of datatime to year-month-day
    \item Eliminate alphanumeric symbols
    \item Uppercase to lowercase
    \item Eliminate words with size less or equal than 3
    \item Add some exceptions
\end{itemize}

\subsection{Visualization}

\begin{itemize}
    \item The date of user account creation
    \item Tweets per day to analyze the increasing number of posts
    \item Cloud of words to analyze the most frequent terms involved per day
\end{itemize}
    
\section{Results}

The next graphics present the results of the experiments and answer many questions to understand the phenomenon over the population.

\subsection{What about the veracity of the posts?}
\begin{figure}[hbpt]
\centerline{\includegraphics [width=0.4\textwidth]{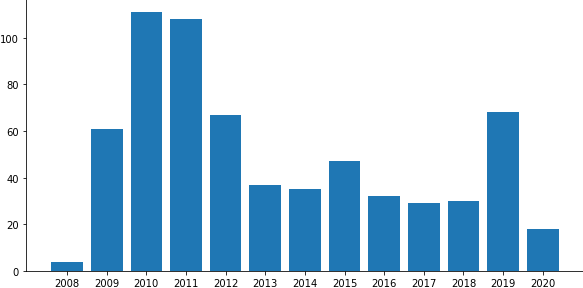}}
\caption{Data User Creation}
\label{fig:duc}
\end{figure}

Nowadays, many users are posting their ideas using Social Networks and there is no control over the veracity of the information. For this reason, one field related to this is the date of the creation of the accounts, this information is presented in Fig. \ref{fig:duc}

Analyzing the previous, a concentration of the dates is around 2010, 2011 then the age of this account is greater than 6 years. So, if fake users wants to post false information, usually the age of the account could be less than 1 year.

\subsection{How often people post and where did they start?}

\begin{figure}[hbpt]
\centerline{\includegraphics [width=0.4\textwidth]{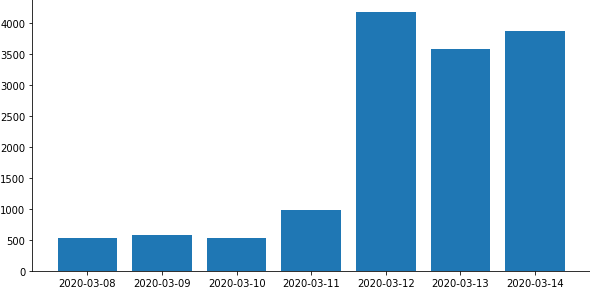}}
\caption{Tweets during the last seven days}
\label{fig:tw7d}
\end{figure}

Considering the window for this analysis was from 29-12-2019 to 14-03-2020, there was an expectation of recovering posts for every day but people was not posting about it during the previous date of 08-03-2020. The graphic Fig. \ref{fig:tw7d} shows an increasing number of post during the last days.

\subsection{What is the people posting about Covid19 symptoms?}

\begin{figure*}[hbpt]
\centerline{\includegraphics [width=0.7\textwidth]{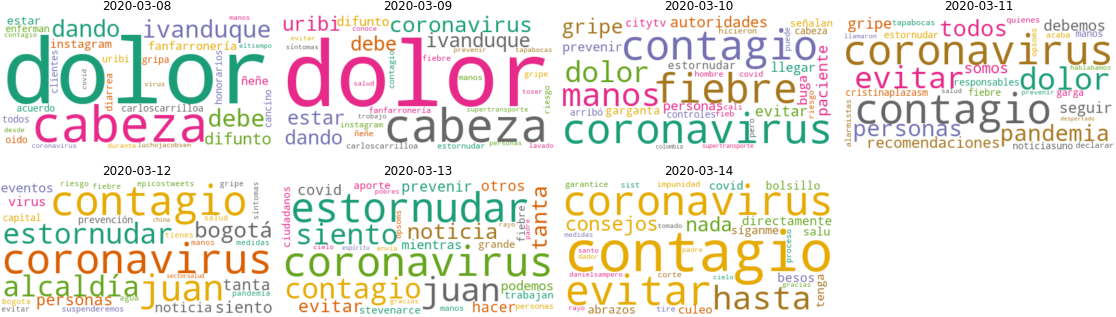}}
\caption{Cloud of Words following the histogram of words}
\label{fig:cwhw}
\end{figure*}

After preprocessing tweets and remove stopwords, the predominant words from 2020-03-08 to 2020-03-14 are: dolor, cabeza, ivanduque, coronavirus, uribi, fiebre, contagio, manos, gripe, evitar, estornudar taken from the cloud of words in Fig.\ref{fig:cwhw}. 

Then then most frequent words introduces topics related symptoms(health), besides the graphic shows interest on politics.

\subsection{How is the progress of covid19 in Colombia?}

\begin{figure}[H]
\centerline{\includegraphics [width=0.4\textwidth]{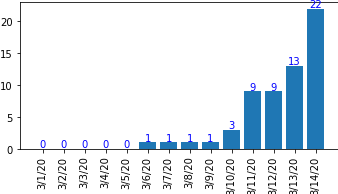}}
\caption{Number of infections in Colombia}
\label{fig:tsco}
\end{figure}

Finally, the image \ref{fig:tsco} shows the actual increase of infection in Colombia from the start of March, and there is a natural correlation between the increasing number of post per day and the number of infections.

This preliminar analysis helps to understand what is happening in the population in Bogota and this data can be useful to analyze others aspects, phenomenon from different approaches: Economy, Sociology, etc.

\section{Conclusions}
A Text Mining approach helps to visualyze what is happening about symptoms of covid19 in Bogota. The relevance of the topic for the people, the increasing number of post, the most relevant terms for day and how the previous ones are naturally correlated to the number of infected people in Colombia.

\section{Recommendations}

API Twitter has a limitation of seven days then if you need to collect data, you must set the range of time. Preprocessing step is necessary because the people posts with any rule on mind. 


\vspace{12pt}
\end{document}